\documentclass{aa}
\usepackage{graphicx}
\usepackage{txfonts}
\usepackage[authoryear]{natbib}
\bibpunct{(}{)}{;}{a}{}{,}
\begin{document}

\title{Wind accretion in the massive X-ray binary \object{4U~2206+54}: abnormally slow wind and a moderately eccentric orbit}

\author{M. Rib\'o\inst{1,2}
\and I. Negueruela \inst{3}
\and P. Blay\inst{4}
\and J. M. Torrej\'on\inst{3}
\and P. Reig\inst{5,6}
}

\offprints{M. Rib\'o}

\institute{DSM/DAPNIA/Service d'Astrophysique, CEA Saclay, B\^at. 709,
L'Orme des Merisiers, 91191 Gif-sur-Yvette, Cedex, France
\and{AIM - Unit\'e Mixte de Recherche - CEA - CNRS - Universit\'e
Paris VII - UMR n$^{\rm o}$7158}\\
\email{mribo@discovery.saclay.cea.fr}
\and Departamento de F\'{\i}sica, Ingenier\'{\i}a de Sistemas y Teor\'{\i}a de la Se\~nal, Escuela Polit\'ecnica Superior, Universitat d'Alacant, Ap. 99, 03080 Alicante, Spain\\
\email{[ignacio;jmt]@dfists.ua.es}
\and Institut de Ci\`encia dels Materials, Universitat de Val\`encia, PO Box 22085, 46071 Valencia, Spain\\
\email{pere.blay@uv.es}
\and IESL, Foundation for Research and Technology, 71110 Heraklion,
Crete, Greece
\and University of Crete, Physics Department, PO Box 2208, 710 03 Heraklion, Crete, Greece\\
\email{pau@physics.uoc.gr}
}

\authorrunning{M. Rib\'o et~al.}

\titlerunning{Wind accretion in the massive X-ray binary \object{4U~2206+54}}

\date{Received / Accepted}

\abstract{Massive X-ray binaries are usually classified by the properties of
the donor star in classical, supergiant and Be X-ray binaries, the main
difference being the mass transfer mechanism between the two components. The
massive X-ray binary \object{4U~2206+54} does not fit in any of these groups,
and deserves a detailed study to understand how the transfer of matter and the
accretion on to the compact object take place. To this end we study an {\it
IUE} spectrum of the donor and obtain a wind terminal velocity ($v_\infty$) of
$\sim$350~km~s$^{-1}$, which is abnormally slow for its spectral type. We also
analyse here more than 9 years of available {\it RXTE}/ASM data. We study the
long-term X-ray variability of the source and find it to be similar to that
observed in the wind-fed supergiant system \object{Vela~X-1}, reinforcing the
idea that \object{4U~2206+54} is also a wind-fed system. We find a
quasi-period decreasing from $\sim$270 to $\sim$130~d, noticed in previous
works but never studied in detail. We discuss possible scenarios for its
origin and conclude that long-term quasi-periodic variations in the mass-loss
rate of the primary are probably driving such variability in the measured
X-ray flux. We obtain an improved orbital period of $P_{\rm
orb}=9.5591\pm0.0007$~d with maximum X-ray flux at MJD~51856.6$\pm$0.1. Our
study of the orbital X-ray variability in the context of wind accretion
suggests a moderate eccentricity around 0.15 for this binary system. Moreover,
the low value of $v_\infty$ solves the long-standing problem of the relatively
high X-ray luminosity for the unevolved nature of the donor,
\object{BD~+53$\degr$2790}, which is probably an O9.5\,V star. We note that
changes in $v_\infty$ and/or the mass-loss rate of the primary alone cannot
explain the different patterns displayed by the orbital X-ray variability. We
finally emphasize that \object{4U~2206+54}, together with \object{LS~5039},
could be part of a new population of wind-fed HMXBs with main sequence donors,
the natural progenitors of supergiant X-ray binaries.
}

\maketitle

\section{Introduction} \label{intro}

High Mass X-ray Binaries (HMXBs) are X-ray sources composed of an early-type
massive star and an accreting compact object, either a neutron star (NS) or a
black hole (BH). Depending on the nature of the companion, HMXBs are
traditionally divided \citep[see][]{corbet86} into three groups: Classical
Massive X-ray binaries, Supergiant X-ray binaries (SXBs) and Be/X-ray binaries
(BeXBs).

Classical Massive X-ray binaries are a very small group (in the Galaxy, only
\object{Cen~X-3}) of very bright ($L_{{\rm X}}\sim10^{38}$~erg~s$^{-1}$)
persistent X-ray sources \citep{xraybinaries95}. They have close orbits, with
short orbital periods, and accretion is believed to occur through localised
Roche-lobe overflow leading to the formation of an accretion disc. Orbits have
circularised and, when the compact object is an NS, it has been spun up to
short spin periods, as a result of angular momentum transfer from the accreted
matter on to the compact object.

In SXBs, the X-ray source is believed to be fed by direct accretion from the
relatively dense wind of an OB supergiant, with little angular momentum
transfer \citep[see][]{waters89}, resulting in moderately high X-ray
luminosities, $L_{{\rm X}}\sim10^{36}$~erg~s$^{-1}$. The orbital periods of
these systems are typically $P_{{\rm orb}}\la 15$~d. Some orbits are almost
circular \citep{corbet02}, but finite eccentricities have been measured for
several systems, some of them rather high (e.g., $e=0.17$ for 4U~1538$-$52;
\citealt{clark00}). The spin periods of systems with NSs are rather long,
$P_{{\rm spin}}> 100$~s.

In BeXRBs, an NS orbits an unevolved OB star surrounded by a dense equatorial
disc. These systems can be bright X-ray transients or persistent low
luminosity $L_{{\rm X}}\sim 10^{34}$~erg~s$^{-1}$ sources. With the exception
of a few peculiar cases like the microquasar \object{LS~I~+61~303}
(\citealt{massi04}, and references therein) all of them appear to be X-ray
pulsars. Their orbital eccentricities range from close to zero to very high,
leading to the hypothesis that different kinds of supernova explosion are
possible \citep{pfahl02a}, and their orbital periods are generally of the
order of some tens of days. Spin periods also cover a wide range and there is
a strong statistical correlation between $P_{{\rm orb}}$ and $P_{{\rm spin}}$
\citep{corbet86}, suggesting effective transfer of angular momentum from the
material accreted.

The peculiar HMXB \object{4U~2206+54} is difficult to place within this
picture. It is a persistent source, with $L_{{\rm
X}}\simeq10^{35}$--$10^{36}$~erg~s$^{-1}$ and variability on timescales of
hours similar to those of wind accreting systems. However, it is one of the
few HMXBs not displaying X-ray pulsations, although there is strong evidence
that the compact object is an NS \citep{torrejon04,blay05}. The X-ray flux is
modulated with a periodicity of $9.568\pm0.004$~d \citep{corbet01}, which can
only be interpreted as the orbital period. The mass donor in
\object{4U~2206+54}, \object{BD~+53$\degr$2790}, is neither a supergiant nor a
Be star, but a peculiar late O-type star, whose spectrum does not admit a
standard spectral classification \citep{negueruela01}. While most criteria
favour an O9.5\,V star, there are some indications of a much heavier mass loss
than expected for that spectral type, such as strong H$\alpha$ emission, a
P-Cygni profile in \ion{He}{ii}~$\lambda$\,4686 and a very strong P-Cygni
profile in the ultraviolet \ion{C}{iv} resonance doublet, suggesting a more
luminous star (\citealt{negueruela01}; \citealt{blay06}). If it has the
luminosity of a normal O9.5\,V star, it is located at a distance of
$\simeq$2.6~kpc (if it were an O9.5\,III star it would be located at
$\simeq$4.8~kpc).

In a recent paper we used {\it INTEGRAL} and VLA data to constrain the nature
of the compact object in \object{4U~2206+54} \citep{blay05}. Based on
existing radio/X-ray correlations for black holes in the low/hard state, we
excluded the black hole scenario. On the other hand, {\it INTEGRAL} is the
third X-ray satellite providing marginal evidence for the presence of a
cyclotron absorption line, leading to an NS with $B=3.6\times10^{12}$~G.
However, two problems persisted in the NS scenario: the lack of X-ray
pulsations and an X-ray luminosity one to two orders of magnitude higher than
expected. 

Here we analyse an ultraviolet spectrum obtained with the {\it International
Ultraviolet Explorer} ({\it IUE}), to better constrain the wind properties of
the donor, and more than 9 years of data from the {\it Rossi X-ray Timing
Explorer} ({\it RXTE}) to derive information on the mass loss from the optical
star and the binary parameters. This work is organised as follows: in
Sect.~\ref{iue} we analyse and model the {\it IUE} spectrum, in
Sect.~\ref{data} we present the {\it RXTE} data, in Sect.~\ref{long} we study
and discuss the long-term X-ray variability, in Sect.~\ref{orbital} we focus
on the orbital X-ray variability, and discuss the long-term wind variability
in Sect.~\ref{windvar}. We stress the existence of a population of wind-fed
HMXBs with main sequence donors in Sect.~\ref{ms} and summarise our
conclusions in Sect.~\ref{conclusions}.

\section{A measure of the wind terminal velocity} \label{iue}

The observed X-ray luminosity of \object{4U~2206+54} is in the range
$\sim$10$^{35}$--10$^{36}$~erg~s$^{-1}$ \citep{blay05}. In contrast, the
expected Bondi-Hoyle accretion luminosity for a canonical NS in a 9.6~d orbit
around a low-luminosity O9.5\,III--V star, with a typically fast wind of
$\sim$1500~km~s$^{-1}$, is of the order of or below 10$^{34}$~erg~s$^{-1}$.
This value is critically influenced by the wind terminal velocity, $v_\infty$.
To obtain a measure of $v_\infty$ for the wind of \object{BD~+53$\degr$2790}
we have analysed the only publicly available high-resolution UV spectrum of
this star, obtained with {\it IUE} on 1990 June 18--19, with a total exposure
time of 20~ks (middle time at JD~2\,448\,061.59). This is the high-dispersion
{\it IUE} spectrum SWP~39112, described in \citet{negueruela01}, but we have
used the new reduction available at the INES\footnote{{\tt
http://ines.vilspa.esa.es/ines/}} database. A heliocentric velocity correction
of 16.37~km~s$^{-1}$ has already been applied to the source spectrum, and we
have further applied a correction of $-$62.7~km~s$^{-1}$ to account for the
radial velocity of \object{BD~+53$\degr$2790} \citep{abt63}. We note that this
value is the average of three measurements spanning 55.0--72.2~km~s$^{-1}$,
obtained from September 1961 to May 1962 and covering different orbital phases
(the precise phase of each data point is uncertain due to the error in the
orbital period and the huge timespan between these observations and the
current ephemeris). Moreover, a preliminary radial velocity curve
\citep{blay06t} shows total relative variations up to $\pm$30~km~s$^{-1}$,
with a relative mean value of $\sim$0$\pm$10~km~s$^{-1}$ around the orbital
phase when the {\it IUE} spectrum was obtained. Therefore, we estimate that
the radial velocity correction is accurate to $\sim$10~km~s$^{-1}$.

\begin{figure*}[t!]
\center
\resizebox{1.0\hsize}{!}{\includegraphics[angle=0]{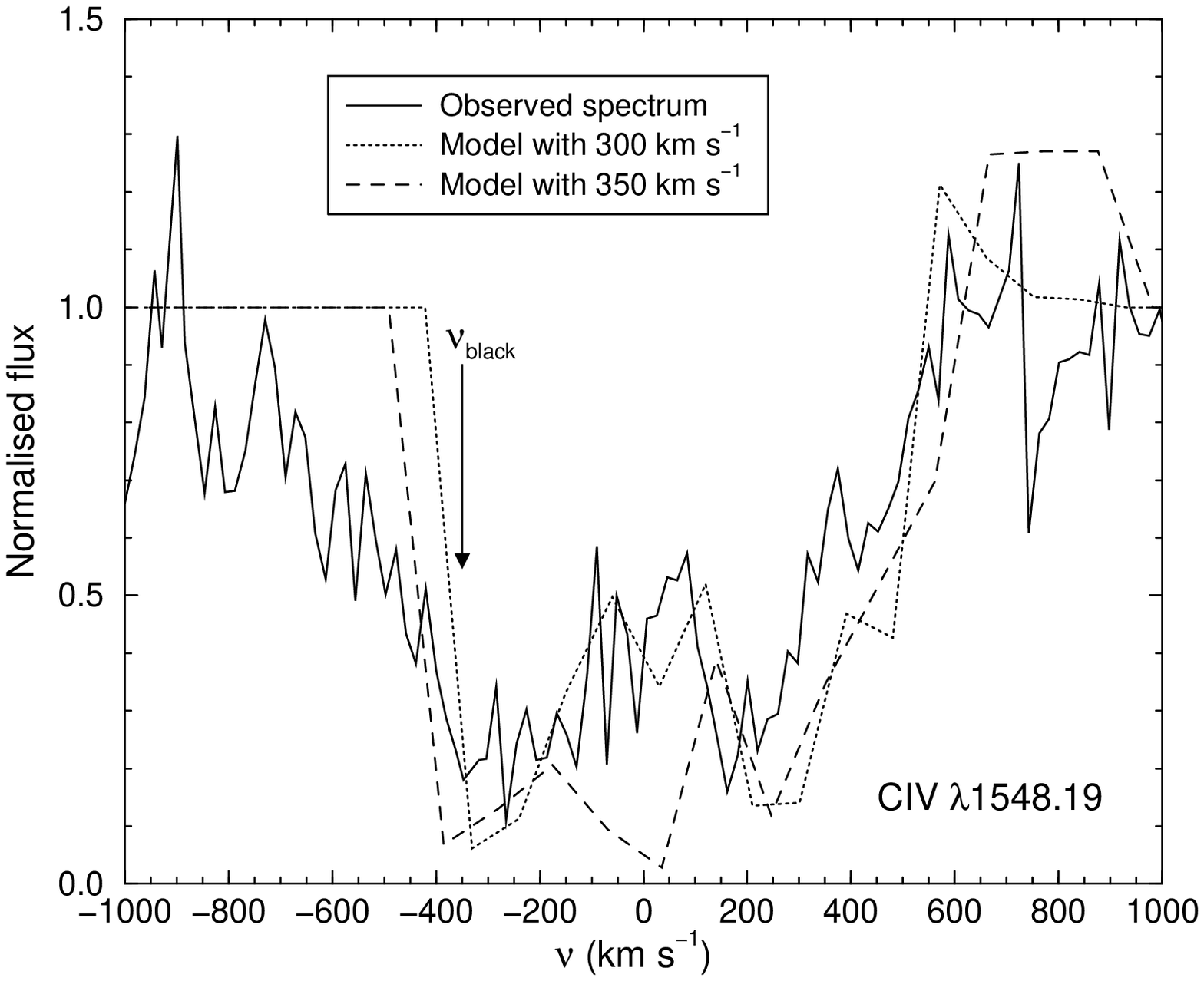}\hspace{3mm}\includegraphics[angle=0]{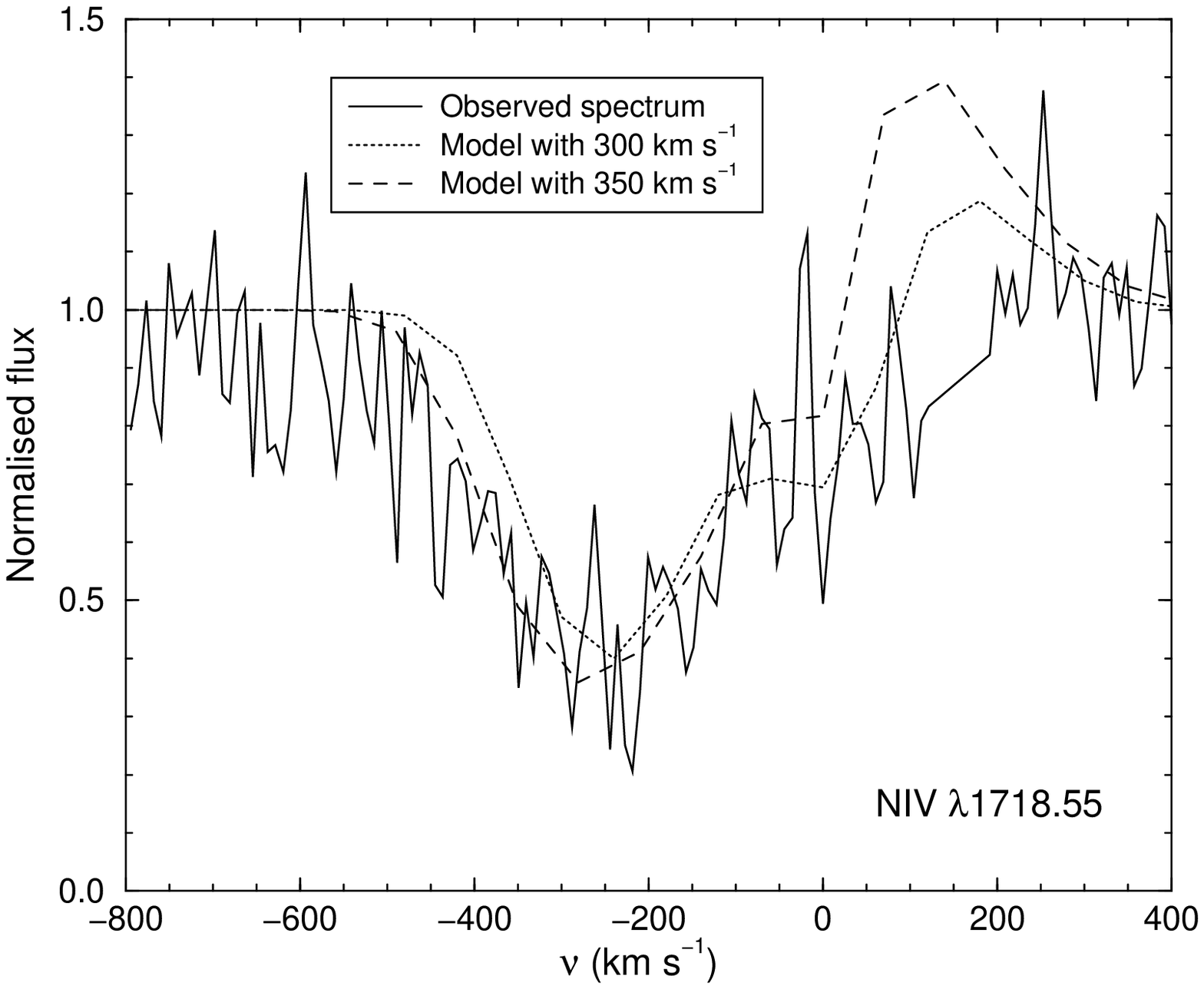}}
\caption{Profiles of the observed UV doublet
\ion{C}{iv}~$\lambda\lambda$~1548.19--1550.76 ({\it left panel}, where
$v=0$~km~s$^{-1}$ corresponds to the blue line of the doublet) and the line
\ion{N}{iv}~$\lambda$\,1718.55 ({\it right panel}) with the theoretical models
for 300 and 350~km~s$^{-1}$ superimposed. The arrow in the left plot indicates
$v_{\rm black}$=350~km~s$^{-1}$. The wind terminal velocity is clearly slower
than 500~km~s$^{-1}$.}
\label{fig:wind_uv}
\end{figure*}

We have followed the SEI (Sobolev with Exact Integration of the transfer
equation) method, as outlined in \citet{lamers87} in order to calculate
theoretical wind line profiles. This method is a modification of the Sobolev
approximation to the stellar wind problem \citep[see][]{sobolev60}. The
procedure followed was to create a grid of theoretical profiles with different
wind and photospheric parameters and then match them all against the observed
profile. The indications from the extensive study of the stellar winds of
\citet{groenewegen89} were followed and the parameters listed there for an
O8\,V star were chosen as initial values to build our grid of models.
Theoretical models could not reproduce the data when using terminal velocities
above 500~km~s$^{-1}$. Terminal velocities in the range 300--350~km~s$^{-1}$
and a turbulent motion with a mean velocity in the range 20--100~km~s$^{-1}$
yielded line profiles which resulted in the best match against the observed
profiles. We show in Fig.~\ref{fig:wind_uv} an example of two UV lines from
\object{BD~+53$\degr$2790} matched against two theoretical profiles,
calculated for 300 and 350~km~s$^{-1}$. For the
\ion{C}{iv}~$\lambda\lambda$\,1548.19--1550.76 doublet an upper limit for the
turbulent velocity of 80~km~s$^{-1}$ was found, while for the
\ion{N}{iv}~$\lambda$\,1718.55 line this limit was found to be
100~km~s$^{-1}$. We note that, due to the presence of many wind features in
this spectral region and the relatively poor signal-to-noise ratio of the {\it
IUE} spectrum, the position of the photospheric continuum is quite uncertain,
and the choice of normalisation parameters can affect our results. However,
the \ion{C}{iv} doublet is saturated, and \cite{prinja90} showed that the
point where saturated wind lines turn upwards towards the continuum level,
namely $v_{\rm black}$, is a good estimator of $v_\infty$. As shown in
Fig.~\ref{fig:wind_uv} left, the \ion{C}{iv} doublet provides $v_{\rm
black}=v_\infty\simeq350$~km~s$^{-1}$, in good agreement with the above
estimates by using the SEI method. Finally, we used the automatic fitting
procedure developed by \cite{georgiev05}, based on genetic algorithms, which 
provides $v_\infty\simeq450$~km~s$^{-1}$ for the \ion{C}{iv} doublet and
$v_\infty\simeq350$~km~s$^{-1}$ for the \ion{N}{iv} line, also in good
agreement with the values quoted above.

Although an accurate measurement of the wind terminal velocity in
\object{BD~+53$\degr$2790} is prevented by the relatively noisy {\it IUE}
spectrum, we stress that only models with $v_\infty<500$~km~s$^{-1}$ could
reproduce the wind line positions and widths. This value is much lower than
the range of 1120--1925~km~s$^{-1}$ for O9\,V stars or the
1275--1990~km~s$^{-1}$ range for O9.5\,III stars \citep{prinja90}. Therefore,
we conclude that the stellar wind of \object{BD~+53$\degr$2790} is abnormally
slow for its spectral type. A similarly slow wind of 400~km~s$^{-1}$ has been
measured in the O9.5\,V extreme fast rotator \object{HD~93521}
(\citealt{prinja90}, excluded from their mean because it had very peculiar
profiles; see also \citealt{howarth93} and \citealt{massa95}). Another O-type
star with an abnormally slow wind, of 510~km~s$^{-1}$ \citep{prinja90}, is
\object{HD~37022} (also known as \object{$\theta^1$~Ori~C}), a spectral
variable in the range O4--7\,V (\citealt{walborn81}; \citealt{smith05}), whose
anomalous properties are interpreted in terms of a misaligned magnetic
rotator. Interestingly, \citet{blay06} have found that
\object{BD~+53$\degr$2790} has a high rotational velocity, $v \sin i =
315\pm70$~km~s$^{-1}$, and have suggested a possible connection with
\object{HD~37022} based on their optical characteristics.

\section{The {\it RXTE}/ASM data} \label{data}

In order to study the long-term X-ray behaviour of \object{4U~2206+54} we have
analysed X-ray data in the energy range 1.3--12.1~keV obtained with the All
Sky Monitor (ASM) on board {\it RXTE}. The {\it RXTE}/ASM data used here spans
from 1996 late February to 2005 late July (from MJD~50135 to MJD~53576),
amounting to a total of 3441 days or 9.42 years (the previous analysis by
\citealt{corbet01} was performed with $\sim$5.5 years of data). Each data
point in the original lightcurve represents the fitted source flux of a 90~s
pointing or `dwell' on the source, with a mean of $\simeq$18.3 dwells per day
in the case of \object{4U~2206+54}. We have also analysed the one-day average
lightcurve of individual {\it RXTE}/ASM dwells (see \citealt{levine96} for
details). This lightcurve contains 3315 flux measurements, with data lacking
only for 126 days (less than 4\% of the total). Since \object{4U~2206+54} is a
weak X-ray source, when there are only a few individual dwells per day the
one-day average flux derived is not very reliable. To avoid spurious points
with large error bars, we have also constructed a one-day average lightcurve
with at least 5 dwells per day, which contains 3050 data points (92\% of the
one-day average data), and another lightcurve with at least 10 dwells per day,
which contains 2608 data points (79\% of the one-day average data). We will
consider all these four lightcurves when performing the timing analysis, and
we will refer to them as DBD (Data By Dwell), ODA (One-Day Averages), 5D-ODA
(5 Dwell ODA) and 10D-ODA (10 Dwell ODA).

\begin{figure*}[t!]
\center
\resizebox{0.8\hsize}{!}{\includegraphics{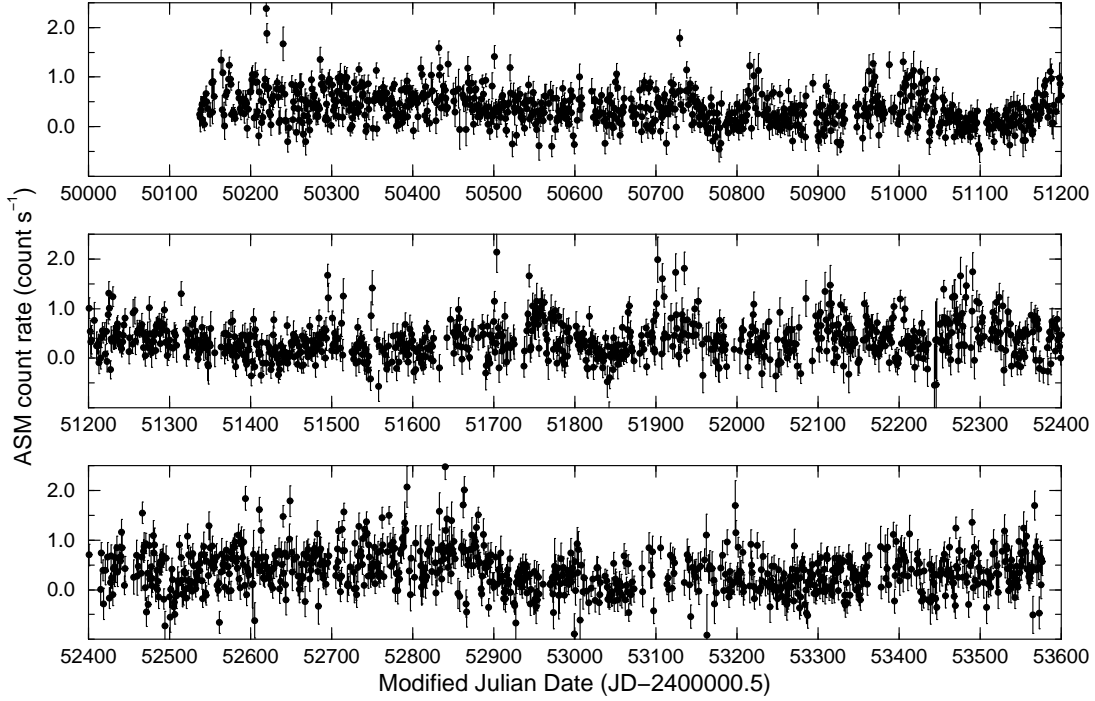}}
\caption{{\it RXTE}/ASM one-day average lightcurve of \object{4U~2206+54}
after removing all the averages containing less than 10 individual dwells
(10D-ODA). Each panel represents approximately 3.3 years of data. Long-term
X-ray flux variations on timescales of hundreds of days are seen.}
\label{asm_lc}
\end{figure*}

The resulting 10D-ODA lightcurve is shown in Fig.~\ref{asm_lc}. The source is
clearly detected during most of the 9.42 year coverage, with a mean count rate
of $0.36^{+0.40}_{-0.31}$ count~s$^{-1}$ (using weights as 1/$\sigma^2$; the
plus and minus standard deviation of the mean have been computed separately
for the points above and below it). Assuming a distance of 2.6~kpc to
\object{4U~2206+54} \citep{blay06}, and taking into account that the average
Crab count rate in the {\it RXTE}/ASM is 75.5 count~s$^{-1}$, we obtain a
weighted mean and standard deviation of the absorbed X-ray luminosity of
$L_{(1.3-12.1~{\rm keV})}\simeq\left(1.4^{+1.5}_{-1.2}\right)\,(d/2.6~{\rm
kpc})^2\times10^{35}$~erg~s$^{-1}$. We note that using a hydrogen column
density of $N_{\rm H}=1.1\times10^{22}$~atoms~cm$^{-2}$ \citep{torrejon04} and
the formalism described in \cite{gallo03}, the unabsorbed luminosity would
only be around 5\% higher.

\section{Long-term X-ray variability} \label{long}

\begin{figure}[t!]
\center
\resizebox{1.0\hsize}{!}{\includegraphics{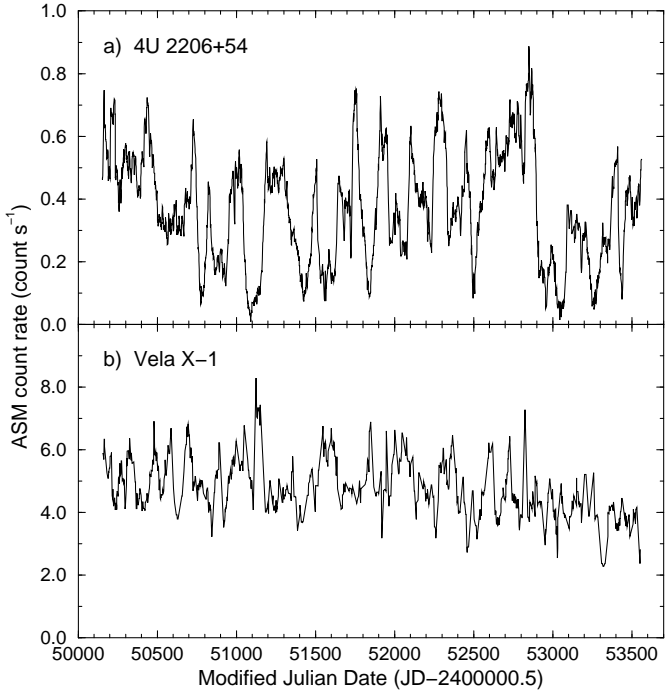}}
\caption{{\bf a)} The data shown in Fig.~\ref{asm_lc} smoothed with a running
window of 30 days of length, allowing a much better visualisation of the
long-term flux variability. {\bf b)} The same procedure applied to the {\it
RXTE}/ASM data of the wind-fed HMXB \object{Vela~X-1} (after excluding
the data taken during its X-ray eclipses). Recurrent peaks and troughs are
seen in both lightcurves.}
\label{asm_lc_s}
\end{figure}

Long-term variability of the mean X-ray flux is clearly seen in the {\it
RXTE}/ASM data, as pointed out by \citet{corbet01}. To display this
variability more clearly, we plot in Fig.~\ref{asm_lc_s}a the same data as in
Fig.~\ref{asm_lc} but after averaging (using weights as 1/$\sigma^2$) all data
points within a running window of 30-day length (corresponding to $\sim$3
times the orbital period). The count rate varies between 0.009 and 0.9
count~s$^{-1}$, with a mean of 0.31 and a standard deviation of 0.17. For
comparison, the mean obtained with adjacent 30~d windows is 0.31, with a
standard deviation of 0.16 and an error of the mean of 0.015, implying that
the variability is real at an 11-$\sigma$ significance. Similar results are
obtained by using running windows of 20 and 10 days, although spurious points
reaching even negative flux values appear, due to the presence of intervals
with few data points in the lightcurve. Therefore, we used the data shown in
Fig.~\ref{asm_lc_s}a to compute the range in luminosities, which turns out to
be $L_{(1.3-12.1~{\rm keV})}\simeq(0.035$--$3.5)\,(d/2.6~{\rm
kpc})^2\times10^{35}$~erg~s$^{-1}$. This confirms the flux variations with a
factor of $\sim$100 on timescales of years noted by \citet{masetti04} when
comparing data from {\it EXOSAT}, {\it RXTE} and {\it BeppoSAX}, but now
obtained with the same satellite and detector (although the value of the
minimum flux is uncertain in this case).

This kind of lightcurve does not show any resemblance to that of any known
Be/X-ray binary, persistent or transient. It is, however, similar to those of
wind-accreting X-ray binaries. As a comparison, we have plotted in
Fig.~\ref{asm_lc_s}b the {\it RXTE}/ASM lightcurve of the wind accretor
\object{Vela~X-1}, averaged and smoothed in exactly the same way as that of
\object{4U~2206+54} (after excluding data taken during X-ray eclipses, which
otherwise translates into a mean reduction in flux of $\sim$1.2
count~s$^{-1}$). \object{Vela X-1} has rather similar orbital parameters to
\object{4U~2206+54}: it contains an NS ($P_{\rm spin}=283$~s) in a low
eccentricity orbit ($e=0.0898$, $P_{\rm orb}=8.9644$~d) around the B0.5Ib star
\object{HD~77581} (see \citealt{quaintrell03}, and references therein). As can
be seen, the lightcurve obtained for \object{Vela~X-1} shows less variability
(a factor $\sim$4) but, like \object{4U~2206+54}, it also experiences
recurrent peaks and troughs, with no clear periodicity (see next subsection).

As noted by \citet{negueruela01} and by \citet{blay06},
\object{BD~+53$\degr$2790} is not a Be star, but a peculiar O9.5\,V star with
a relatively strong stellar wind. Therefore the comparison to the SXBs is
fully justified, as the mass-loss mechanism in the mass donors is likely to be
identical, i.e., a radiative stellar wind (see \citealt{kudritzki00}, and
references therein). The main difference between the two kinds of systems is
the higher luminosity class of the mass donors in SXBs. This higher luminosity
class will result in a higher X-ray luminosity through two effects: 1) a
higher mass-loss rate at the base of the wind and 2) a larger radius for the
mass donor, which will place the NS closer to the surface of the OB star and
hence in the region where the wind is denser and slower (although we have
shown in Sect.~\ref{iue} that the wind of \object{BD~+53$\degr$2790} is
abnormally slow). Apart from this difference, all available evidence supports
the idea that \object{4U~2206+54} is a wind accretor, not fundamentally
different from the SXBs.

We have inspected if there is any trend of hardening or softening of the
spectrum as the flux increases on these long-term timescales. We have done so
by computing all possible hardness ratios that can be constructed with the
three energy bands of {\it RXTE}/ASM data (1.3--3.0, 3.0--5.0, and
5.0--12.1~keV). We have worked with the 10D-ODA lightcurve before and after
smoothing. All the results obtained are compatible, within errors, with a
constant hardness as the flux increases, although we emphasize that the poor
statistics are a strong limitation of the analysis of {\it RXTE}/ASM data of
\object{4U~2206+54}.

\subsection{Timing analysis} \label{timinglong}

\begin{figure}[t!]
\center
\resizebox{1.0\hsize}{!}{\includegraphics{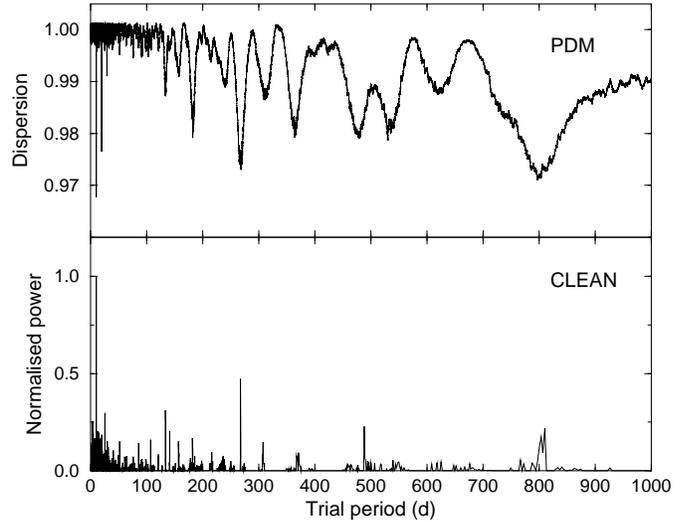}}

\caption{Periodograms of the 10D-ODA lightcurve of \object{4U~2206+54}
obtained by using the PDM ({\it top}) and CLEAN ({\it bottom}) algorithms.
Apart from the $\sim$9.6~d orbital period, a trial period around 270~d is
clearly detected by both methods.}
\label{pdm_clean}
\end{figure}

We have searched for periodic signals in all lightcurves using standard
techniques like the Phase Dispersion Minimisation (PDM,
\citealt{stellingwerf78}) and the CLEAN algorithm \citep{roberts87}. The
periodograms obtained for the 10D-ODA lightcurve between 2 and 1000~d are
shown in Fig.~\ref{pdm_clean}. As can be seen, the orbital period of
$\sim$9.56~d is clearly detected with both methods (with two subharmonics in
the case of PDM). On the other hand, significant signal is detected
simultaneously with both methods around trial periods of $\sim$133, notably
$\sim$267, $\sim$488 and $\sim$800~d. PDM also detects a 1-year signal, with a
harmonic and a subharmonic, which is probably the result of our window
function, since it is not detected by CLEAN. Apart from the orbital period,
these trial periods are also clearly detected when analysing the smoothed data
shown in Fig.~\ref{asm_lc_s}a. We note that a peak at a frequency of
$\simeq4\times$10$^{-3}$~d$^{-1}$ (corresponding to a period of 250~d) was
already present in the power spectrum presented by \citet{corbet01} in their
figure~2, and also noticed by \citet{masetti04}, who questioned if it could be
a superorbital periodicity (see \citealt{clarkson03a,clarkson03b}, for recent
discussions on the topic).

To investigate further the long-term variability, we have split the 10D-ODA
lightcurve in 2 equal data sets spanning 1720~d each, and then re-applied the
PDM and CLEAN algorithms. The PDM results show a period in the range
$\sim$255--265~d in the first part of the lightcurve, with a harmonic and two
broad subharmonics. For the second part of the lightcurve there are minima
around 180 and 415~d, and no significant signal around 260~d. The output of
the CLEAN algorithm reveals similar differences: 256~d, its harmonic and less
significant peaks for the first data set, and different peaks at 90, 180~d,
and many less significant peaks in the second data set. All this indicates
that we are not dealing with a periodic signal of $\sim$260~d, but with a
quasi-periodic one, present in the first part of the lightcurve but not in the
second one, and that the flux varies on timescales of hundreds of days. This
quasi-period of $\sim$260~d can be easily seen as alternative local maxima and
minima in Fig.~\ref{asm_lc_s}a during the first $\sim$5 years of data (similar
to the data analysed by \citealt{corbet01}). We note that very similar results
to the ones discussed above are obtained when analysing the DBD, ODA and
5D-ODA lightcurves.

\begin{figure}[t!]
\center
\resizebox{1.0\hsize}{!}{\includegraphics{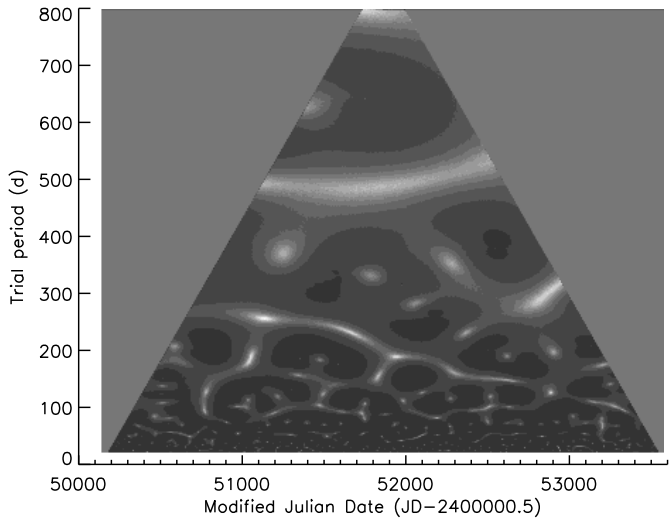}}
\caption{Wavelet amplitude map of the 10D-ODA lightcurve of
\object{4U~2206+54}. A quasi-period decreases from $\sim$270 to $\sim$130~d
during the timespan of the data.}
\label{wav}
\end{figure}

A better understanding of how this quasi-period changes with time can be
achieved with a two-dimensional time-period method. For this purpose, we have
used the method presented by \citet{szatmary94}. We compute $W(f,\tau)$, which
we will call the wavelet amplitude. This value will be high if the signal
contains a cycle frequency $f$ at the time $\tau$, and low otherwise. We have
considered trial periods from 20 to 800~d, with a resolution of 1~d, and a
total of 1720 times of analysis have been taken into account, which
corresponds to one point every 2~d. We show in Fig.~\ref{wav} the wavelet
amplitude map obtained, where white coloured areas mean significant trial
periods and black areas non-significant ones, while the two grey triangular
areas have not been explored due to the presence of severe border effects (see
\citealt{ribo01} for a detailed discussion of this issue). One can see the
variation of the long-term $\sim$200~d quasi-period over time, as well as the
signal around 500~d and marginally at around 800~d. 

\begin{figure}[t!]
\center
\resizebox{1.0\hsize}{!}{\includegraphics{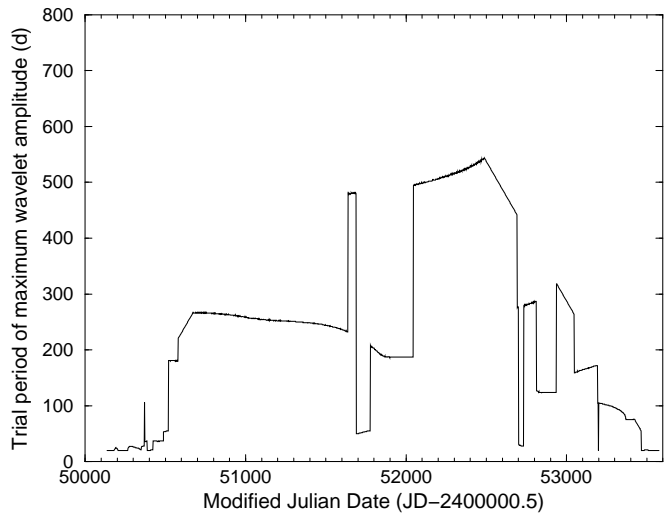}}
\caption{Maxima of the wavelet amplitude map shown in Fig.~\ref{wav} versus
time for the 10D-ODA lightcurve of \object{4U~2206+54}. The decrease of the
quasi-period, with superimposed jumps, is clearly seen.}
\label{wavmax}
\end{figure}

In Fig.~\ref{wavmax} we plot, for any considered time $\tau$ in Modified
Julian Date, the trial period that displays the maximum wavelet amplitude. A
decreasing quasi-period is present in the data, and excluding very short trial
periods found at the limits of the data set and eventual jumps to the
$\sim$500~d signal, it varies from $\sim$270~d at the beginning of the
observations to $\sim$130~d at the end. Very similar results are obtained when
performing the same kind of analysis but using the smoothed data shown in
Fig.~\ref{asm_lc_s}a.

In contrast, a similar analysis of the \object{Vela~X-1} data with PDM, CLEAN
and the wavelet based method reveals no such kind of long-term quasi-periodic
variability in this supergiant binary system.

\subsection{Superorbital quasi-period or wind variability?}

Superorbital periods with $P_{\rm sup}$ in the range 30--240~d and $P_{\rm
sup}/P_{\rm orb}$ values in the range 5--22000 have been found in a group of
around 15 X-ray binaries (see \citealt{wijers99}; \citealt{ogilvie01}, and
references therein). These superorbital periods have often been explained as a
precession of the accretion disc due to warping induced as a consequence of
illumination from the central source. In the case of \object{4U~2206+54}, if
the quasi-periodic signal were a superorbital period, the $P_{\rm sup}/P_{\rm
orb}$ ratio would decrease from $\sim$28 to $\sim$14 during the time interval
covered by the {\it RXTE}/ASM data. The variability of this quasi-period
appears much higher than in other systems. Interaction of modes could be
invoked to explain such variations, and even the jumps to higher values of the
period \citep{ogilvie01}. However, as discussed in \citet{torrejon04}, there
is no evidence for the existence of an accretion disc in \object{4U~2206+54},
and conversely, there are strong reasons to think that the X-ray emission
originates because of direct accretion on to the surface of an NS from the
stellar wind, as discussed above and in Sect.~\ref{orbital}. Therefore, we are
inclined to think that the long-term X-ray variability we see is due, as
suggested by \citet{masetti04}, to variations in the wind of the primary.

Erratic variability on timescales of hours has been detected with {\it RXTE}
and {\it BeppoSAX}, and attributed to wind density inhomogeneities. While in
the first case \citet{negueruela01} reported an increase of the 5--10/2--5~keV
hardness ratio with increasing 2.5--30~keV intensity, in the second one
\citet{masetti04} report a constant behaviour for the same hardness ratio with
increasing 2--10~keV intensity. This later result suggests that variations in
the accretion rate due to variations in the wind density do not change the
slope of the soft X-ray spectrum. We have used the same scheme of hardness
ratio versus intensity as in the later case with the {\it RXTE}/ASM and found
the same behaviour. Despite the large errors present in these data, this would
be consistent with the hypothesis that long-term X-ray variability is due to a
changing wind of the primary.

Interestingly, correlated variations between a variable emission component in
H$\alpha$ and X-ray flux have been observed on timescales of years in the
wind-fed HMXB system \object{LS~5039}/\object{RX~J1826.2$-$1450}
(\citealt{reig03}; \citealt{mcswain04}; \citealt{bosch05}). By analogy, the
long-term variations in the X-ray flux from \object{4U~2206+54} might be due
to changes in the wind of the mass donor \object{BD~+53$\degr$2790}.
Stochastic variability in the wind may easily explain variations in the X-ray
flux on timescales of hundreds of days with no periodicity. It is, however,
much less obvious what physical mechanism could result in the quasi-periodic
variability observed in \object{4U~2206+54}.

\object{BD~+53$\degr$2790} is too early to fit within any known category of
pulsating stars -- no $\beta$ Cep star is known with a spectral type earlier
than B0 \citep{lesh73,tian03}. Moreover, the timescales of variability are
much longer than in any B-type pulsator of any kind \citep{waelkens98}. The
possibility that non-radial pulsation activity is present in the Oef star
\object{BD~+60$\degr$2522} has been suggested to explain variability on
time-scales of hours \citep{rauw03}, but the variability in
\object{BD~+53$\degr$2790} has a timescale more typical of Mira variables.

Although clearly \object{BD~+53$\degr$2790} cannot share a physical mechanism
with Mira variables, there must be a physical reason driving the long-term
quasi-periodic variability that we have detected. Long-term
(quasi-)periodicities have been observed in a few other peculiar O-type stars,
such as \object{HD~108} \citep{naze01}. The recent discovery of a 538-d
recurrence in the spectral changes of the Of?p star \object{HD 191612}
\citep{walborn04} represents another interesting example of long-term
periodicity of unclear origin. While for \object{HD 191612}, the possibility
that a binary companion in a very eccentric orbit drives the changes is
tenable, the quasi-periodic changes in \object{BD~+53$\degr$2790} cannot be
explained by a hypothetical third body in the system.

Whatever the physical driver, if the changes in the X-ray flux are associated
with variability in the mass loss of \object{BD~+53$\degr$2790}, the average
X-ray flux of \object{4U~2206+54} may be a good tracer of the recent wind
history of the donor. Taking into account the observed correlation between
H$\alpha$ emission and X-ray flux in \object{LS~5039}, it may be worth
considering the possibility that, at least in well detached HMXBs with wind
accretion, X-ray monitoring might provide information on the long-term
evolution of wind characteristics.

However, the existence of chaotic short-term variability on timescales from
minutes to days in the X-ray flux from \object{4U~2206+54}
\citep{negueruela01,torrejon04,masetti04} prevents the comparison between
available optical spectroscopy and X-ray flux measurements taken even a few
hours apart. Clearly, a detailed multiwavelength campaign has to be undertaken
to be able to fully interpret the data in each energy domain.

\section{Orbital X-ray variability} \label{orbital}

\citet{corbet00} performed a preliminary analysis of $\sim$4.5 years of {\it
RXTE}/ASM data of \object{4U~2206+54} (after correction from a wrong position
by $\sim$0.5$\degr$ in the survey catalogue). They found a quasi-sinusoidal
modulation with a period of 9.570$\pm$0.004~d, and suggested it could be due
to orbital motion. Later on, \citet{corbet01} performed a detailed analysis of
$\sim$5.5 years of {\it RXTE}/ASM data and refined this value to
9.568$\pm$0.004~d. The folded lightcurve appeared again quasi-sinusoidal, and
they derived an epoch of maximum X-ray flux at MJD~51006.1$\pm$0.2.

\subsection{Timing analysis} \label{timingorbital}

As already explained in Sect.~\ref{timinglong}, we have performed a detailed
timing analysis of the currently available 9.42 years of {\it RXTE}/ASM data.
The orbital period is significantly detected in all data sets and using both
the PDM and CLEAN algorithms. Since the PDM is an epoch folding method and
there is clearly a great deal of long-term variability, we have preferred to
trust the results obtained with CLEAN. We note that, since the folded
lightcurve is quasi-sinusoidal, CLEAN is well suited for this analysis
\citep[see, e.g.,][]{otazu02,otazu04}. The ODA and 5D-ODA lightcurves contain
spurious data, and since the results obtained with CLEAN may be influenced by
them (because weights are not used), we have not considered these cases. For
the DBD and 10D-ODA cases, the results obtained are 9.5589$\pm$0.0005 and
9.5593$\pm$0.0005~d, respectively. Fitting a cosine function to the DBD
lightcurve provides a period of $9.5591\pm0.0007$~d and an epoch of maximum
X-ray flux at MJD~51856.6$\pm$0.1. This will be the ephemeris considered
hereafter. We note, however, that splitting the DBD lightcurve in two halves
provides cosine fits with periods of $9.5694\pm0.0014$~d and
$9.5520\pm0.0026$~d, respectively, with a difference of 0.017~d, much higher
than the error of the fit quoted above. This is probably due to the
superposition of long-term variability.

\begin{figure}[t!]
\center
\resizebox{1.0\hsize}{!}{\includegraphics{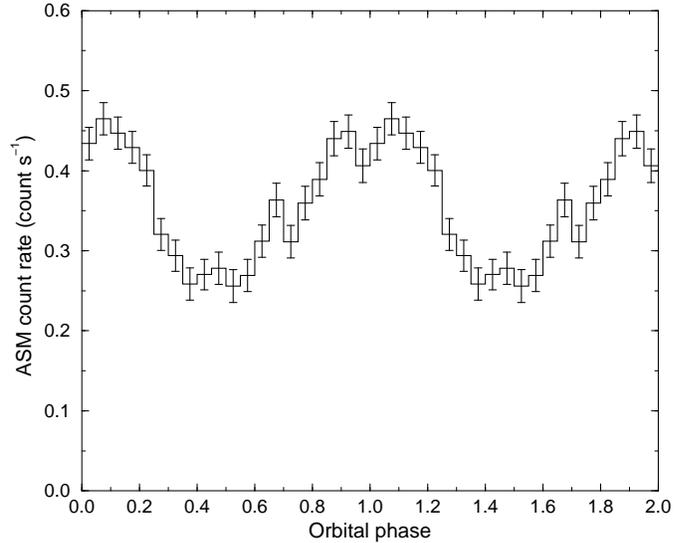}}
\caption{DBD lightcurve of \object{4U~2206+54} folded using $P_{\rm
orb}$=9.5591~d and $t_0$=MJD~51856.6. Error bars represent the error of the
weighted mean in each of the 20 bin per period, and not the standard
deviation. Two orbital periods are shown for clarity. A quasi-sinusoidal
pattern is evident.}
\label{f:orbital}
\end{figure}

\begin{table*}[t!]
\begin{center}
\caption[]{Computed orbital variability of $L_{\rm acc}$ for NSs (with $M_{\rm
X}$=1.4~$M_\odot$ and $R_{\rm X}$=10~km) accreting from the wind of different
donors in eccentric orbits. The values for O9.5\,V and O9.5\,III donors
(\object{4U~2206+54}) have been computed for different wind terminal
velocities. As a comparison we also show the results for \object{Vela~X-1}
considering an inclination of 80\degr\ \citep{quaintrell03}. The observed {\it
RXTE}/ASM luminosities are around 1/3 of $<L_{\rm acc}>$.}
\label{ratio}
\begin{tabular}{@{}l@{~~~~}c@{~~~~}c@{~~~~}c@{~~~~}c@{~~~~}c@{~~~~}c@{~~~~}c@{~~~~}c@{~~~}c@{~}c@{~}c@{}}
\hline \hline \noalign{\smallskip}
Spectral type & $M_{\rm opt}$ & $R_{\rm opt}$ & $\dot{M}_{\rm opt}$   & $v_{\infty}$  & $\beta$ & $P_{\rm orb}$ & $e$ & $L_{\rm acc~max}$ & $L_{\rm acc~min}$ & Ratio in $L_{\rm acc}$ & $<L_{\rm acc}>$\\
              & ($M_\odot$)   & ($R_\odot$)   & ($M_\odot$~yr$^{-1}$) & (km~s$^{-1}$) &         & (d)           &     & (erg~s$^{-1}$)  & (erg~s$^{-1}$)  &                      & (erg~s$^{-1}$)\\
\noalign{\smallskip} \hline \noalign{\smallskip}
O9.5\,V                       & 16.0 & ~~7.3 & 3.0$\times$10$^{-8}$ & ~~350 & 0.8 & 9.5591 &    0.15 & ~~5.53$\times$10$^{35}$ & ~~2.95$\times$10$^{35}$ & 1.87 & ~~4.06$\times$10$^{35}$\\
      &      &       &                                              & ~~500 & 0.8 &        &         & ~~2.19$\times$10$^{35}$ & ~~1.12$\times$10$^{35}$ & 1.96 & ~~1.57$\times$10$^{35}$\\
      &      &       &                                              &  1000 & 0.8 &        &         & ~~0.22$\times$10$^{35}$ & ~~0.10$\times$10$^{35}$ & 2.10 & ~~0.15$\times$10$^{35}$\\
O9.5\,III                     & 20.8 &  13.3 & 2.0$\times$10$^{-7}$ & ~~350 & 0.8 & 9.5591 &    0.15 & 33.4~~$\times$10$^{35}$ & 17.9~~$\times$10$^{35}$ & 1.86 & 24.7~~$\times$10$^{35}$\\
      &      &       &                                              & ~~500 & 0.8 &        &         & 15.2~~$\times$10$^{35}$ & ~~7.54$\times$10$^{35}$ & 2.02 & 10.8~~$\times$10$^{35}$\\
      &      &       &                                              &  1000 & 0.8 &        &         & ~~1.85$\times$10$^{35}$ & ~~0.79$\times$10$^{35}$ & 2.34 & ~~1.21$\times$10$^{35}$\\
\noalign{\smallskip} \hline \noalign{\smallskip}
\object{Vela X-1} (80\degr, 1.96~$M_\odot$) & 24.2 & 28.0 & 1.0$\times$10$^{-6}$ & 1100 & 0.8 & 8.9644 & 0.0898 & ~~5.47$\times$10$^{36}$ & ~~2.52$\times$10$^{36}$ & 2.17 & ~~3.75$\times$10$^{36}$\\
\noalign{\smallskip} \hline
\end{tabular}
\end{center}
\end{table*}

We show in Fig.~\ref{f:orbital} the DBD lightcurve folded by using these
ephemerides. We have weighted the data according to 1/$\sigma^2$ in each one
of the 20 bins per period used, where $\sigma$ is the error of each individual
dwell. As can be seen, the modulation is quasi-sinusoidal, but showing a
slightly slower rise and a faster decay. This behaviour is unaffected by small
changes in the $P_{\rm orb}$ used or by small shifts in $t_0$ to average the
data in a different way, indicating that it might be real. Nevertheless we
must be cautious about this issue, since the long-term variability can affect
the folded lightcurve in a noticeable way. On the other hand, the local
minimum around maximum (phases 0.95--1.00) is washed out when using slightly
different values for $P_{\rm orb}$ and/or $t_0$, which makes it more
uncertain.

As a further check of the orbital variability, and in an attempt to avoid the
influence of the long-term variability, we have analysed the data in the
following way. First of all, we have only considered the DBD data points 
falling on intervals of time when the average flux shown in
Fig.~\ref{asm_lc_s}a was in the range 0.25--0.55 count~s$^{-1}$, to avoid
intervals of poor statistics and intervals of too much activity of the source.
This corresponds approximately to 57\% of the original DBD data. After that,
we subtracted the average flux from the DBD data, to remove the remaining
long-term variability. Finally, we analysed the resulting data set as
previously done. CLEAN provides an orbital period of 9.5587$\pm$0.0005~d,
compatible with the one reported above. The folded lightcurve obtained by
using this filtered data set and the quoted period is very similar to the one
shown in Fig.~\ref{f:orbital} (but with a mean count rate of 0). The local
minimum around phases 0.95--1.00 is also present, although again with a low
significance. Regarding the orbital period, we prefer the former value,
$9.5591\pm0.0007$~d, because it was obtained with the original dataset, which
contained about 2 times the number of flux measurements used in the latter
case.

\subsection{Wind accretion in an eccentric orbit?} \label{wind}

The minimum and maximum {\it RXTE}/ASM count rates in Fig.~\ref{f:orbital} are
0.26 and 0.46 count~s$^{-1}$, respectively, implying a ratio between them of
1.8. Translating the observed count rates into absorbed X-ray luminosities, we
obtain an orbital variability covering the range $L_{(1.3-12.1~{\rm
keV})}\simeq(1.0$--$1.8)\,(d/2.6~{\rm kpc})^2\times10^{35}$~erg~s$^{-1}$. This
degree of variability is expected in a wind-accreting system with a low or
moderate eccentricity. In order to explore whether the observed variability
can provide constraints on system parameters, we have used a beta-law with
spherical symmetry to model the wind of the donor, computed the position and
velocity of the compact object in an eccentric orbit around it, and obtained
the luminosity due to accretion by using a Bondi-Hoyle accretion model
\citep{bondi44,bondi52}. A detailed explanation of the method is given in
\cite{reig03}. A comparison between the equation we have used to estimate the
X-ray luminosity due to accretion, $L_{\rm acc} = G M_{\rm X} \dot{M}_{\rm
acc}/R_{\rm X}$, and that of the commonly used formalism by \cite{lamers76},
where $L_{\rm acc} = \zeta \dot{M}_{\rm acc} c^2$, reveals that the efficiency
factor for the conversion of accreted matter to X-ray flux is $\zeta = G
M_{\rm X}/R_{\rm X} c^2 = 1.48 (M_{\rm X}/M_\odot)/(R_{\rm X}/{\rm km})$. 

Since this is a very simple accretion model, we used \object{Vela~X-1} to
check its validity. Folding the {\it RXTE}/ASM data of the source with its
orbital period, we find maximum and minimum count rates of 6.4 and
$\la$3.2~count~s$^{-1}$ (the presence of the X-ray eclipse close to the
minimum X-ray flux prevents an accurate estimate in this latter case), thus
providing a luminosity ratio $\ga$2.0. Assuming a distance to the source of
1.9~kpc \citep{sadakane85}, these count rates translate into absorbed X-ray
luminosities of $1.3\times10^{36}$ and $\la 0.7\times10^{36}$~erg~s$^{-1}$,
with an average of $<L_{(1.3-12.1~{\rm
keV})}>\simeq1.0\times10^{36}$~erg~s$^{-1}$. On the other hand, by using $\log
L/L_\odot = 5.53$ \citep{sadakane85} and the relationship by \cite{howarth89},
we obtain a mass-loss rate of
$\dot{M}\simeq1\times10^{-6}$~$M_\odot$~yr$^{-1}$. For the wind terminal
velocity we use the value $v_{\infty}= 1100$~km~s$^{-1}$, obtained from {\it
IUE} spectra \citep{prinja90}. Hereafter we will use $\beta=0.8$ for the
exponent of the wind law. The orbital period and the eccentricity are those of
\cite{quaintrell03}, while we have considered their intermediate case of
$i$=80\degr, which provides $M_{\rm opt}=24.2~M_\odot$, $R_{\rm
opt}=28.0~M_\odot$, and $M_{\rm X}=1.96~M_\odot$ (with the adopted values,
$\zeta\simeq0.3$). The expected accretion luminosity values are quoted in the
last row of Table~\ref{ratio}. As can be seen, we obtain a ratio in $L_{\rm
acc}$ of 2.2, compatible with the $\ga$2.0 ratio obtained with the {\it
RXTE}/ASM data. On the other hand we find $<L_{(1.3-12.1~{\rm
keV})}>\simeq<L_{\rm acc}>/~4$, since not all the accretion luminosity is
released in the soft X-ray band. Indeed, from {\it BeppoSAX} data of
\object{Vela~X-1} by \cite{orlandini98}, we found that between 1/3--1/4 of the
total X-ray luminosity between 2 and 100~keV would be emitted in the {\it
RXTE}/ASM energy range. Therefore, the accretion luminosities obtained through
our model yield a ratio of maximum to minimum luminosity compatible with the
{\it RXTE}/ASM data, and absolute values a factor of 3 to 4 above them, in
agreement with observations.

\begin{figure}[t!]
\center
\resizebox{1.0\hsize}{!}{\includegraphics{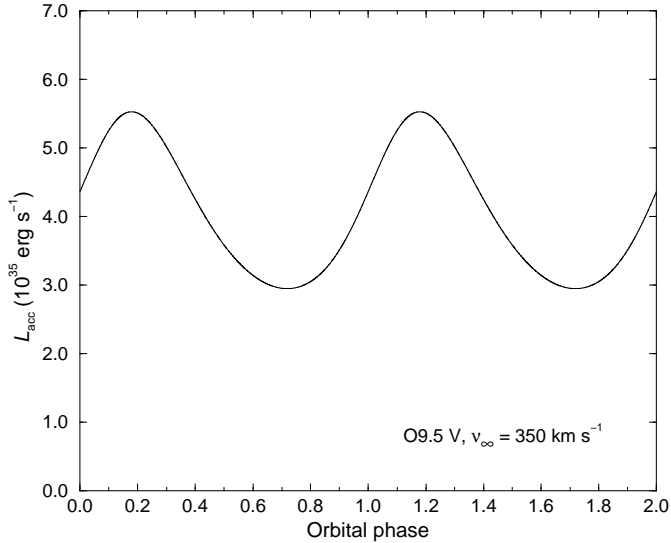}}
\caption{Computed variability of the accretion luminosity from the compact
object in \object{4U~2206+54}, assuming a neutron star with $M_{\rm
X}=1.4~M_\odot$ and $R_{\rm X}=10$~km, using a spherically symmetric wind and a Bondi-Hoyle accretion model. The variability is due to the orbital
motion in an eccentric orbit with $e=0.15$ around an O9.5\,V donor with
$v_{\infty}=350$~km~s$^{-1}$. All parameters are quoted in the first row of
Table~\ref{ratio}. A quasi-sinusoidal pattern, similar to the observed one,
shown in Fig.~\ref{f:orbital}, is obtained. The observed soft X-ray luminosity
is $\sim$1/3 the accretion luminosity. Phase 0 corresponds to periastron, and
the maximum of accretion luminosity is displaced due to the low wind
velocity.}
\label{accretion}
\end{figure}

In the case of \object{4U~2206+54} we have used $M_{\rm X}=1.4~M_\odot$ and
$R_{\rm X}=10$~km for the NS (which provides $\zeta\simeq0.2$). We have run
models with the parameters corresponding to both an O9.5\,V star and an
O9.5\,III star. In the first case, we considered $M_{\rm opt}=16.0~M_\odot$
and $R_{\rm opt}=7.3~R_\odot$ \citep{martins05}, with $\dot{M}_{\rm
opt}=3\times10^{-8}$~$M_\odot$~yr$^{-1}$ (by using $\log L/L_\odot = 4.65$
from \citealt{martins05} and the relationship by \citealt{howarth89}). In the
second case, and using the same references, we have $M_{\rm opt}=20.8~M_\odot$
and $R_{\rm opt}=13.3~R_\odot$, with $\dot{M}_{\rm opt}=2 \times
10^{-7}M_\odot$~yr$^{-1}$. We have tried different eccentricities and found
that we can approximately reproduce the quasi-sinusoidal shape of the folded
lightcurve shown in Fig.~\ref{f:orbital} by using $e=0.15$, as can be seen in
Fig.~\ref{accretion} for the O9.5\,V star considered above and using
$v_{\infty}=350$~km~s$^{-1}$. We note that in this last figure phase 0
corresponds to periastron, and the maximum accretion luminosity takes place at
phase $\sim$0.2 because of the low wind terminal velocity (i.e., if this model
were correct, periastron would be at phase $\sim$0.8 in Fig.~\ref{f:orbital}).
The results of our simulations using an eccentricity of 0.15, for both the
main sequence and giant cases, and different values for the wind terminal
velocity are quoted in Table~\ref{ratio}. As can be seen in the main sequence
case for $v_{\infty}=350$~km~s$^{-1}$, we are able to reproduce the 1.8
maximum to minimum X-ray luminosity ratio and obtain an accretion luminosity
$\la$3 times the observed one (of $<L_{(1.3-12.1~{\rm
keV})}>=1.4\times10^{35}$~erg~s$^{-1}$), similar to the factor of $\sim$2
between the 4--150 to 4--12~keV luminosities \citep{blay05}. However, for
higher wind velocities we underestimate the average X-ray luminosity,
consistent with the low-speed wind observed by {\it IUE} (see
Sect.~\ref{iue}). In the giant case the system would be at 4.8~kpc, and the
average X-ray luminosity $<L_{(1.3-12.1~{\rm
keV})}>=4.8\times10^{35}$~erg~s$^{-1}$. Therefore we should expect average
accretion luminosities between 15 and $20\times10^{35}$~erg~s$^{-1}$, which
can be achieved with $v_{\infty}\simeq400$~km~s$^{-1}$, again compatible with
the {\it IUE} data. We caution, nevertheless, that the wind terminal velocity
has been derived with UV data taken several years before the X-ray flux
measurements form {\it RXTE}/ASM.

It is important to note that, for the unevolved nature of
\object{BD~+53$\degr$2790} and the orbital parameters of the system, the low
value of the wind velocity allows \object{4U~2206+54} to have an unexpectedly
high X-ray luminosity even if containing an NS of 1.4~$M_\odot$ as the
accreting compact object. This solves the main problem in accepting the NS
hypothesis, the other one being the lack of X-ray pulses that can be explained
by simple geometrical effects \citep{blay05}. Finally, from the absence of
X-ray eclipses, and for $e=0.15$, we derive an upper limit on the orbit
inclination of 82.5\degr\ in the O9.5\,V case, and 77.5\degr\ in the O9.5\,III
case.

\section{On the long-term wind variability} \label{windvar}

\begin{figure}[t!]
\center
\resizebox{1.0\hsize}{!}{\includegraphics{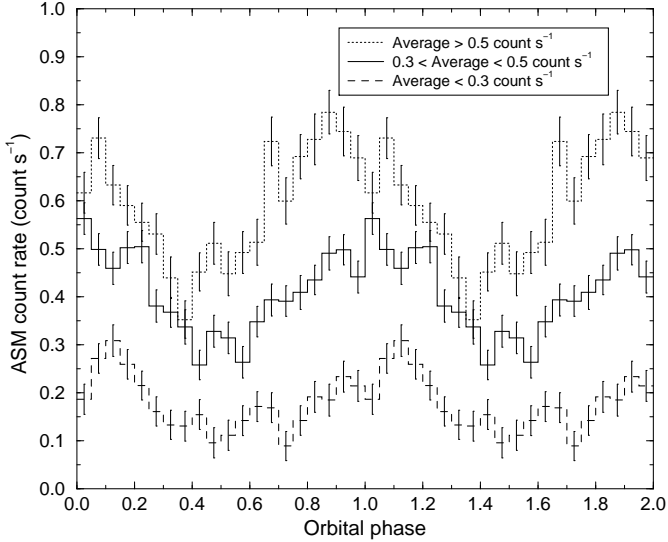}}
\caption{The same as Fig.~\ref{f:orbital} but splitting the data in 3
different 30-d-average flux levels: high, middle, and low. The middle case is
similar to the total shown in Fig.~\ref{f:orbital}, with a broad maximum
centered at phase 0. When the average flux is high, the maximum X-ray flux
happens at phase $\sim$0.85, while for the low flux case the maximum is around
phase 0.10. This behaviour is contrary to what would be expected if the
average X-ray flux increases because of a decrease in the wind terminal
velocity.}
\label{f:orbital_hml}
\end{figure}

There are two simple ways to explain the long-term changes in the X-ray flux
from \object{4U~2206+54}: changes in the mass-loss rate of the donor and/or
changes in the wind terminal velocity. Assuming a constant mass-loss rate of
$\dot{M}_{\rm opt}=3\times10^{-8}$~$M_\odot$~yr$^{-1}$, wind terminal velocity
changes between 200 and 1100~km~s$^{-1}$ should be invoked to provide
accretion luminosities in the range $10^{34}$--$10^{36}$~erg~s$^{-1}$,
necessary to explain the observed long-term X-ray variability discussed in
Sect.~\ref{long}, of $L_{(1.3-12.1~{\rm
keV})}\simeq(0.035$--$3.5)\,(d/2.6~{\rm kpc})^2\times10^{35}$~erg~s$^{-1}$.
These velocity changes would lead to different orbital variability patterns:
for the lower values of $v_\infty$ we would have a higher X-ray flux with a
maximum peaking at a phase 0.28 after periastron, while for the higher values
of $v_\infty$ the lower X-ray flux would have its maximum around phase 0.06
after periastron. To check if this is the case, we have split the DBD data in
three intervals according to whether the 30-day averages shown in
Fig.~\ref{asm_lc_s}a are in the following ranges: $>0.5$~count~s$^{-1}$ (24\%
of data), $>0.3$ and $<0.5$~count~s$^{-1}$ (38\% of data),
$<0.3$~count~s$^{-1}$ (38\% of data). We have subsequently folded these data
with the orbital period, as done in Fig.~\ref{f:orbital}. We show the results
in Fig.~\ref{f:orbital_hml}. For higher count rates the maximum X-ray flux
takes place earlier than for lower count rates, contrary to what would be
expected if only changes in $v_\infty$ drove the changes in X-ray luminosity. 

If only changes in the mass-loss rate of the primary are invoked, then for
$v_{\infty}=350$~km~s$^{-1}$ we would need dramatic long-term variations in
the range $\dot{M}_{\rm
opt}=7\times10^{-8}$--$7\times10^{-10}$~$M_\odot$~yr$^{-1}$. Moreover, a
change in the mass-loss rate does not modify the phase of the X-ray luminosity
maximum. Despite the relatively noisy {\it RXTE}/ASM data of
\object{4U~2206+54}, we can say that the scenario is not so simple.

A general explanation of the various regimes of wind accretion on to a
magnetised neutron star can be found in \cite{stella86}. In our case, the
accretion radius $r_{\rm acc}=2GM_{\rm X}/v_{\rm rel}^2$ \citep{waters89}
varies along the eccentric orbit, and takes values in the range
2--3$\times$10$^{11}$~cm for the $v_{\infty}=350$~km~s$^{-1}$ case
($\sim$1.5$\times$10$^{11}$~cm for $v_{\infty}=500$~km~s$^{-1}$, and
$\sim$0.5$\times$10$^{11}$~cm for $v_{\infty}=1000$~km~s$^{-1}$). This is one
order of magnitude smaller than the separation of both objects at periastron
passage. On the other hand, by using a surface magnetic field of
$B_0=3.6\times10^{12}$~G \citep{blay05} and the dipole magnetic field formula
\citep[$B(r)=B_0 (R_{\rm X}/r)^3$;][]{waters89} we can compute the magnetic
radius along the orbit, given by $r_{\rm mag}=(B_0^2 R_{\rm X}^6/ 8\pi\rho
v_{\rm rel}^2)^{1/6}$, which takes values in the range
0.6--1.0$\times$10$^{10}$~cm for all wind values considered. Indeed,
considering the cases quoted in Table~\ref{ratio} and all possible orbital
phases, the accretion radius is always between 5 and 40 times larger than the
magnetospheric radius. Finally, since we do not know the spin period of the
NS, we cannot compute the corotation radius. However, if we want to avoid
centrifugal inhibition of accretion, also known as the propeller mechanism
\citep[see][]{stella86}, we must have $r_{\rm cor}=(GM_{\rm X}P_{\rm
spin}^2/4\pi^2)^{1/3}>r_{\rm mag}$. Therefore, to have direct wind accretion
in \object{4U~2206+54}, the spin period has to be longer than 470~s. We have
computed the magnetospheric radius as a function of the orbital phase, and it
is at its minimum just before periastron and shifted in phase $\sim$0.3 before
the maximum of accretion luminosity. Therefore, we suggest that the behaviour
shown in Fig.~\ref{f:orbital_hml} could be the result of enhanced accretion
when the magnetospheric radius is minimum, during epochs of high mass-loss
rates of the primary.

In addition, geometric variations in the X-ray irradiation of the wind of the
primary can lead to changes in the ionization states of the wind species that
we see, resulting in variations of the measured $v_{\infty}$
\citep{hatchett77}. This effect has been observed in \object{Vela~X-1}
\citep{kaper93} and in \object{4U~1700$-$37} \citep{iping04}. Moreover, for
higher X-ray irradiation one has a higher ionization and a slower wind, thus a
higher accretion rate and X-ray luminosity. The eccentricity of
\object{4U~2206+54} could lead to additional significant changes in the amount
of X-ray irradiation of the primary wind along the orbit, and thus in
$v_{\infty}$, providing a different orbital variability pattern from the one
obtained with a simple Bondi-Hoyle model. There are certain epochs when the
{\it RXTE}/ASM maximum to minimum orbital flux ratio appears to be a factor of
$\sim$5. This can be seen in data with relatively good signal-to-noise ratios
after averaging with a 2-day running window. An example of this variability is
shown in Figure~4 of \cite{corbet01}. If this behaviour were only due to
accretion variability in an eccentric orbit, a value of $e\simeq0.4$ would be
needed. However, these factor of $\sim$5 flux variations are not seen in the
average folded lightcurve. Even when considering the split data in
Fig.~\ref{f:orbital_hml}, variations with a factor in the range 2--3 are seen.
Clearly, a radial velocity curve is needed to constrain the eccentricity of
\object{4U~2206+54} and allow a better modeling of the data.

\section{HMXBs with main sequence donors} \label{ms}

\object{4U~2206+54} has been known since the early days of X-ray astrophysics
\citep{giacconi72}, although a complete understanding of its accreting
properties has only been possible after analysing an {\it IUE} spectrum. The
abnormally slow wind of the donor in this binary system,
$\sim$350~km~s$^{-1}$, results in a relatively high X-ray luminosity,
$10^{35}$--$10^{36}$~erg~s$^{-1}$, for its relatively wide orbit, $\sim$9.6~d,
which has allowed its detection in all X-ray surveys. In addition, the system
is relatively nearby, 2.6~kpc \citep{blay06}, and does not suffer an
extremely high absorption, $N_{\rm H}=1.1\times10^{22}$~atoms~cm$^{-2}$
\citep{torrejon04}. Similar systems containing main sequence donors with
normal fast winds would have one to two orders of magnitude lower X-ray
luminosities, preventing their detection in existing X-ray surveys.

The only exception is the nearby (2.5~kpc) X-ray binary system
\object{LS~5039} (see \citealt{casares05}, and references therein). The donor
is an O6.5\,V((f)) star \citep{clark01} displaying a fast wind of
2440~km~s$^{-1}$ \citep{mcswain04}, but the system is present in the {\it
ROSAT} All Sky Bright Source Catalog (\citealt{voges99}; \citealt{motch97}),
because of its close orbit, $\sim$3.9~d \citep{casares05}, and relatively low
absorption, $N_{\rm H}=0.7\times10^{22}$~atoms~cm$^{-2}$ \citep{martocchia05}.
We note that the X-ray luminosity of this system, $\simeq10^{34}$~erg~s$^{-1}$
\citep{bosch05}, is only a small fraction, a factor of 1/80, of the accretion
luminosity \citep{casares05}, making its detection even more difficult.

In this context, sensitive pointed observations of the Galactic centre with
{\it Chandra} revealed the existence of a population of $\sim$1000
low-luminosity hard X-ray sources \citep{wang02,muno03}. Although a
substantial fraction of them could be wind-fed accreting NSs in HMXBs with
main sequence donors \citep{pfahl02b}, their nature is still under debate (see
\citealt{bandyopadhyay05} and \citealt{laycock05}).

On the other hand, available all-sky (or Galactic-plane) X-ray surveys are not
sensitive and/or hard enough to detect sources of this kind a few kpc away
from us, where interstellar absorption probably plays a crucial role.
Therefore, the existence of a population of wind-fed HMXBs with main sequence
donors (and $P_{\rm spin}>100$~s in the case of NSs to avoid the propeller
effect), which would be the natural progenitors of SXBs, could be unveiled if
new sensitive all-sky surveys with energies above $\sim$5~keV are performed
(like the planned {\it ROSITA}\footnote{\tt
http://www.rssd.esa.int/index.php?project=Rosita} and {\it EXIST}\footnote{\tt
http://exist.gsfc.nasa.gov/} surveys). Their study could have fundamental
consequences for models of evolution and population synthesis of binary
systems.

\section{Conclusions} \label{conclusions}

After a study of an {\it IUE} spectrum of \object{BD~+53$\degr$2790} and
$\sim$9.4 years of {\it RXTE}/ASM data of \object{4U~2206+54}, we conclude
that:

\begin{enumerate}

\item The ultraviolet spectrum reveals that, in mid June 1990, the O9.5\,V
star \object{BD~+53$\degr$2790} had a wind terminal velocity of
$\sim$350~km~s$^{-1}$, abnormally slow for its spectral type, but similar to
the ones measured in two peculiar fast rotators.

\item An improved orbital period of $9.5591\pm0.0007$~d was obtained, to be
compared with the slightly longer period of 9.568$\pm$0.004~d reported
previously by \citet{corbet01} using the first $\sim5.5$ years of {\it
RXTE}/ASM data.

\item Long-term X-ray flux variability on timescales of hundreds of days is
present in the data, compatible with a quasi-period that decreases from
$\sim$270 to $\sim$130~d during the course of the {\it RXTE}/ASM monitoring
between February 1996 and July 2005. We conclude that this reflects changes in
the wind of the donor, whose quasi-periodic nature remains puzzling.

\item Using a Bondi-Hoyle accretion model, a spherically symmetric wind with
$v_{\infty}=350$~km~s$^{-1}$ and an eccentric orbit with $e\simeq0.15$, we are
able to reproduce quite well the average X-ray luminosity variations of the
source with the orbital period, as well as the absolute X-ray luminosity of
the system, which no longer poses problems in accepting a (non-pulsating) NS
as the compact object in this binary system.

\item The different patterns of the orbital X-ray variability for different
average X-ray fluxes indicates that the long-term X-ray variability cannot be
explained by variations in $v_{\infty}$ and/or $\dot{M}_{\rm opt}$ alone. Due
to the orbital eccentricity we expect changes in the magnetospheric radius and
perhaps changes in $v_{\infty}$ caused by variations in the ionization states
in the wind of the primary due to changes in the X-ray irradiation. These
parameters could vary as well on long-term scales for different mass-loss
rates of the primary. We suggest that they might play a role in explaining the
different patterns of orbital X-ray variability.

\item If the magnetic field of the NS is $B_0=3.6\times10^{12}$~G
\citep{blay05}, its spin period has to be longer than 470~s to allow direct
wind accretion.

\item Long-term coordinated observations in the optical and X-rays could
confirm the proposed variability of the wind, by a correlation between X-ray
flux and H$\alpha$ excess.

\item Observations are underway to obtain the radial velocity curve of
\object{4U~2206+54} and constrain the eccentricity of the system, which is
needed to properly model the observed orbital X-ray variability of the source.

\item The nearby X-ray binaries \object{4U~2206+54} and \object{LS~5039} are
the only two known wind-fed HMXBs with main sequence donors. We suggest that
more sensitive and harder X-ray surveys than the available ones could unveil a
new population of objects of this kind, which are the natural progenitors of
supergiant X-ray binaries.

\end{enumerate}

\begin{acknowledgements}

Based on quick-look results provided by the {\it RXTE}/ASM team.
We thank the anonymous referee for useful comments.
We thank A. Herrero for useful suggestions when analysing the {\it IUE} spectrum and L. Georgiev and X. Hern\'andez for kindly supplying their genetic algorithm code. 
We thank S. Chaty for useful comments on a draft version of the manuscript.
This research is supported by the Spanish Ministerio de Educaci\'on y Ciencia (former Ministerio de Ciencia y Tecnolog\'{\i}a) through grants AYA2001-3092, ESP-2002-04124-C03-02, ESP-2002-04124-C03-03 and AYA2004-07171-C02-01, partially funded by the European Regional Development Fund (ERDF/FEDER). 
M.R. acknowledges financial support from the French Space Agency (CNES)
and by a Marie Curie Fellowship of the European Community programme
Improving Human Potential under contract number HPMF-CT-2002-02053.
I.N. is a researcher of the programme {\em Ram\'on y Cajal}, funded by the Spanish Ministerio de Educaci\'on y Ciencia and the University of Alicante, with partial support from the Generalitat Valenciana and the European Regional Development Fund (ERDF/FEDER).
P.B. acknowledges support by the Spanish Ministerio de Educaci\'on y Ciencia through grant ESP-2002-04124-C03-02.
This research has made use of the NASA Astrophysics Data System Abstract
Service and of the SIMBAD database, operated at the CDS, Strasbourg, France.

\end{acknowledgements}

\end{document}